\documentclass[sn-basic]{sn-jnl}

\usepackage{graphicx}%
\usepackage{multirow}%
\usepackage{amsmath,amssymb,amsfonts}%
\usepackage{amsthm}%
\usepackage{mathrsfs}%
\usepackage[title]{appendix}%
\usepackage{xcolor}%
\usepackage{textcomp}%
\usepackage{manyfoot}%
\usepackage{booktabs}%
\usepackage{algorithm}%
\usepackage{algorithmicx}%
\usepackage{algpseudocode}%
\usepackage{listings}%
\usepackage{cite}%

\begin{document}

\title[Article Title]{\textbf{Optical Optimization of A Multi-Slit Extreme Ultraviolet Spectrograph for Global Solar Corona Diagnostics}}


\author[1,2]{\fnm{Yufei} \sur{Feng}}

\author*[1,2,3,4]{\fnm{Xianyong} \sur{Bai}}\email{xybai@bao.ac.cn}

\author[1,3]{\fnm{Sifan} \sur{Guo}}

\author[5]{\fnm{Hui} \sur{Tian}}

\author[5]{\fnm{Lami} \sur{Chan}}

\author[1,2,3]{\fnm{Yuanyong} \sur{Deng}}

\author[1,3]{\fnm{Qi} \sur{Yang}}

\author[1,2,3]{\fnm{Wei} \sur{Duan}}

\author[1,2]{\fnm{Xiaoming} \sur{Zhu}}

\author[1,2,3]{\fnm{Xiao} \sur{Yang}}

\author[1,2,3]{\fnm{Zhiwei} \sur{Feng}}

\author[1,2,3]{\fnm{Zhiyong} \sur{Zhang}}

\affil*[1]{\orgdiv{National Astronomical Observatories}, \orgname{Chinese Academy of Sciences}, \city{Beijing}, \postcode{1001101}, \country{China}}

\affil[2]{\orgdiv{Key Laboratory of Solar Activity and Space Weather, National Space Science Center}, \orgname{Chinese Academy of Sciences}, \city{Beijing}, \postcode{100190}, \country{China}}

\affil[3]{\orgdiv{School of Astronomy and Space Sciences}, \orgname{University of Chinese Academy of Sciences}, \city{Beijing}, \postcode{100049}, \country{China}}

\affil[4]{\orgdiv{Institute for Frontiers in Astronomy and Astrophysics}, \orgname{Beijing Normal University}, \city{Beijing}, \postcode{102206}, \country{China}}

\affil[5]{\orgdiv{School of Earth and Space Sciences}, \orgname{Peking University}, \city{Beijing}, \postcode{100871}, \country{China}}


\abstract{The spatial-temporal evolution of coronal plasma parameters of the solar outer atmosphere at global scales, derived from solar full-disk imaging spectroscopic observation in the extreme-ultraviolet band, is critical for understanding and forecasting solar eruptions. We propose a multi-slits extreme ultraviolet imaging spectrograph for global coronal diagnostics with high cadence and present the preliminary instrument designs for the wavelength range from 18.3 to 19.8 nm. The instrument takes a comprehensive approach to obtain global coronal spatial and spectral information, improve the detected cadence and avoid overlapping. We first describe the relationship between optical properties and structural parameters, especially the relationship between the overlapping and the number of slits, and give a general multi-slits extreme-ultraviolet imaging spectrograph design process. The multilayer structure is optimized to enhance the effective areas in the observation band. Five distantly-separated slits are set to divide the entire solar field of view, which increase the cadence for raster scanning the solar disk by 5 times relative to a single slit. The spectral resolving power of the optical system with an aperture diameter of 150 mm are optimized to be greater than 1461. The spatial resolution along the slits direction and the scanning direction are about $4.4^{\prime\prime}$ and $6.86^{\prime\prime}$, respectively. The Al/Mo/B$_4$C multilayer structure is optimized and the peak effective area is about 1.60 cm$^2$ at 19.3 nm with a full width at half maximum of about 1.3 nm. The cadence to finish full-disk raster scan is about 5 minutes. Finally, the instrument performance is evaluated by an end-to-end calculation of the system photon budget and a simulation of the observational image and spectra. Our investigation shows that this approach is promising for global coronal plasma diagnostics.}

\keywords{Sun, Corona, Extreme ultraviolet imaging spectroscopy, Full-disk}



\maketitle

\section{Introduction}
\label{sec:intro}
Coronal mass ejections (CMEs) and flares are two of the most violent phenomena in the solar system that can cause disastrous space weather events. Up to now, the physical mechanisms that initiate and drive CMEs and flares, are not fully understood. Currently, it is widely accepted that they are mainly generated in the corona, i.e., the outermost layer of the solar atmosphere. Obtaining global maps of coronal plasma parameters including density, velocity and line width, with a high cadence will significantly improve our understanding of CMEs and flares \citep[e.g.,][]{Tian2021UpflowsIT, delZanna2018SolarUA, Young2021FuturePF}. In addition, such global maps are important for identifying the source regions of solar wind \citep[e.g.,][]{Tian2021UpflowsIT, Brooks2015}, which provides critical constraints for helisopheric models.\par

Extreme ultraviolet (EUV) spectroscopic observations provide a way to realize rapid and accurate measurements of global distributions of coronal plasma parameters. \citeauthor{Urra2023}, \citeyear{Urra2023} claimed that a EUV instrument providing consistent full-disk sampling of plasma properties at the dynamic timescales of solar eruptive phenomena will revolutionize the field by adding the spectral diagnostics. The following parameters should be considered for such instruments: \textbf{(1)} full-disk field of view (FOV) to monitor the global distributions of various coronal features such as active region (AR) loops, coronal holes (CHs),flare, CMEs and filaments, etc.; \textbf{(2)} moderate spatial resolution of about 6 arcseconds or better to resolve EUV bright points, coronal plumes, and AR loops; \textbf{(3)} cadence of several minutes to capture the evolution of the CMEs and flares; \textbf{(4)} covering a narrow spectral band that contains a series of strong coronal lines for diagnosing density and velocity. In other words, an EUV imaging spectrograph that has the ability to rapidly obtain coronal imaging and spectral information with a full-disk FOV is needed.\par

At present, there are four main types of EUV spectrographs used to capture coronal imaging and spectral information.\par

The first one is a single-slit imaging spectroscopy, for example, the Solar Ultraviolet Measurements of Emitted Radiation (SUMER) onboard the Solar and Heliospheric Observatory (SOHO) launched in 1995 \citep{Wilhelm1995SUMERS, Domingo1995TheSM}, the EUV Imaging Spectrograph (EIS) onboard the Hinode mission in 2006 \citep{Culhane2007TheEI, Mariska2007HinodeEI, Korendyke2006OpticsAM} and the Spectral Imaging of the Coronal Environment (SPICE) of the Solar Orbiter mission in 2020 \citep{Anderson2019TheSO}. The solar EUV image is focused by the primary mirror, and partially selected through the slit, dispersed by the grating, and the spectrum of slit is recorded at different positions of the detector depending on the wavelength. Finally, two-dimensional images and spectral information are obtained by scanning the primary mirror step by step. It takes hours to sample a region with a typical size of an active region in this manner, greatly limiting the investigation of the evolution of dynamic events at global scales. In addition, there is a specialized version, exemplified by the Rapid Acquisition Imaging Spectrograph Experiment (RAISE) conducted by the United States Sounding Rocket in 2016 \citep{Laurent2016TheRA}. For the strong radiation lines (Ly $\alpha$ at 121.6 nm), the instrument can observe small-scale multithermal dynamics in AR and the quiet-Sun (QS) loops with its 3 s raster cadence. But the method is not suitable for weaker target lines as the exposure time will be much longer, and scanning the whole solar disk still takes a few hours. \par

The second type is a slitless imaging spectroscopy, such as the S082A payload on \textit{Skylab} in the 1970s \citep{Tousey1977ExtremeUS}, the Multi-Order Solar EUV Spectrograph (MOSES) on a sounding rocket experiment in 2006 \citep{Fox2011SnapshotIS}, the slitless spectrograph of the COronal Spectroscopic Imager in the EUV (COSIE-S) proposed in 2019 \citep{Winebarger2018UnfoldingOS, Golub2020EUVIA} and  the slitless spectrograph of The EUV CME and Coronal Connectivity Observatory (ECCCO-S) approved by NASA in 2023 \citep{Reeves2022ECCCO}. The system consists of only one concave grating and detector, simultaneously recording the $0^{th}$ order broadband image and the mixed spectral image after $\pm1^{st}$ order dispersion (while COSIE records only the spectra of the $2^{nd}$ order of the blazed grating). This kind of instrument has a high cadence, but there is serious overlap from solar images of adjacent spectral lines, increased difficulty in the inversion of corona physical parameters. In addition, the slot detection of EIS instrument, which obtain solar images of large areas in bright solar emission lines with a single exposure, is more like a detected mode of slitless spectrograph. \par

The third type is the multi-slit imaging spectroscopy, for example, the Multi-slit Solar Explorer (MUSE), a MIDEX mission approved by NASA in 2019 \citep{DePontieu2019TheMA, DePontieu2021ProbingTP, Cheung2021ProbingTP}. This approach can be considered as an upgraded version of the first type, featuring multiple closely-separated parallel slits instead of a single slit. Each slit selects a slice on the solar disk. Therefore, the cadence for scanning the same region is reduced several times compared to single-slit imaging spectroscopy. MUSE does not cover a full-disk FOV, as it focused on high spatial resolution (about $0.4^{\prime\prime}$) observation with a FOV of $170^{\prime\prime}\times170^{\prime\prime}$.\par

In addition, a new type of EUV integral field spectrograph (IFS) based on image slicers has emerged, for example, the Spectral Imaging of the Solar Atmosphere (SISA) proposed for the SPARK mission concept \citep{PPR:PPR766453}. The image slicer, which is the most crucial component of the system, divides the solar disk into numerous small field units. It then transmits them to different gratings, allowing for the capture of both imaging and spectral information simultaneously on the detector. The image-slicer-based IFS has also been proposed for spectroscopic observations of solar flares at other wavelengths, including the Lyman Alpha Solar Spicule Observatory (LASSO) for the band of 1205.0 \AA--1220.0 \AA ~\citep{Chamberlin2016AnIF}. Additionally, a Integral Field Unit (IFU) based on image slicers has successfully employed in the GREGOR Infrared Spectrograph on the GREGOR Solar Telescope \citep{Tagle2022}. A machined image slicer integral field unit (MISI) is also being developed for the Diffraction-Limited near-IR Spectropolarimeter (DL-NIRSP) of the Daniel K. Inouye Solar Telescope (DKIST) \citep{Lin2022MISI, Takashi2023IFU, Tetsu2024IFU}. But the manufacturing technology of image slicers for the EUV are currently under development at Technology Readiness Level (TRL) 4. Hence it needs a long time to realize the EUV solar observation with IFS.\par

Obviously, none of the exiting and proposed instrument could adequately address all the detection requirements for solar full-disk spectral diagnostics, particularly the need for a combination of a large FOV, negligible spectral overlapping and high cadence.\par

In this paper we present the preliminary design of a new extreme-ultraviolet imaging spectrograph for global coronal diagnostics with high cadence. Several distantly-separated slits, dividing the full-disk FOV into multiple parts, are designed to carry out the raster scan of the whole solar disk within five minutes. In Section \ref{sec:2}, we describe the requirements and principle of the instrument. Optical system design and optimization of effective area are presented in Section \ref{sec:3}. The theoretical performance of the proposed instrument is evaluated in Section \ref{sec:4}. Finally, we summarize our results in Section \ref{sec:5}. 

\section{Main Instrumental Parameters}
\label{sec:2}

Taking full-disk solar spectral diagnostics at EUV wavelength would: (1) monitor and improve our understanding of highly dynamic solar eruption, e.g., flares and CMEs; (2) derive both of the line-of-sight and plane-of-the-sky velocity of CMEs to better investigating the propagation of CMEs in the interplanetary space and forecasting the arrival time of CME at the earth;  (3) identify the source regions of solar wind. \citeauthor{chen2024global}, \citeyear{chen2024global} described the preliminary consideration for the key instrument parameters. Here we summarize the main results and use them as the input for the instrument design. A field of view of $40^\prime \times 50^\prime$ obtained by rastering 200 steps of $3^{\prime\prime}$ with 5 slits is helpful for monitoring solar activities in the corona. The spectral range of 18.3--19.8 nm is required to measure coronal electron densities (Fe~{\sc{xii}} 195.12/186.89 \AA) and temperatures (Fe~{\sc{viii}}, Fe~{\sc{x}}, Fe~{\sc{xi}}, Fe~{\sc{xii}}) and to investigate the relationship between the evolution of these parameters and solar eruptions. A spatial resolution better than $6^{\prime\prime}$ is required to resolve EUV bright points, coronal plumes, and AR loops. Spectral resolving power, i.e., the central wavelength  $\lambda_0$ divided by the half-height full width $\Delta\lambda_0$, must be greater than 500 to resolve the main diagnostic lines and measure the line-of-sight velocity of a CME \citep{Xu2022SunasastarSO, Yang2022, Lu2023}. A temporal resolution of less than 5 minutes is required for global corona detection to capture the dynamics of CME. The tentatively chosen specifications for the instrument are presented in Table \ref{table:1}. \par

\begin{table}
	\centering
	\caption{Main parameters for the instrument}      
	\label{table:1}      
	\begin{tabular}{cc}   
		\hline  
		Characteristic & Performance Requirement \\   
		\hline                     
		Raster FOV $\Phi$/ $\prime$ & $40^{\prime} \times 50^{\prime}$ in 200 raster steps of $3^{\prime\prime}$ with 5 slits \\
		Temperature ranges/ (log T/K) & 5.65--6.20 \\
		Spectral range $\Delta\lambda$/ nm & 18.3--19.8 (at the center wavelength of 19 nm) \\
		\multirow{2}{*}{Characteristic spectral line} &	Fe~{\sc{viii}} 185.21 \AA, Fe~{\sc{x}} 184.54 \AA, Fe~{\sc{xii}} 188.22 \AA, \\
		& Fe~{\sc{xii}} 186.89 \AA, Fe~{\sc{xii}} 193.51 \AA and Fe~{\sc{xii}} 195.12 \AA \\
		Spatial resolution/ $\prime\prime$ & \textbf{$\sim$ 6} \\
		Spectral resolving power $\lambda_0$/ $\Delta\lambda_0$ & $>$500 ($\lambda_0$ = 19 nm) \\
		Temporal resolution/ s & $\le$ 300 \\
		\hline                     
	\end{tabular}
\end{table}

\begin{figure}
	\centering
	\includegraphics[width=0.95\linewidth]{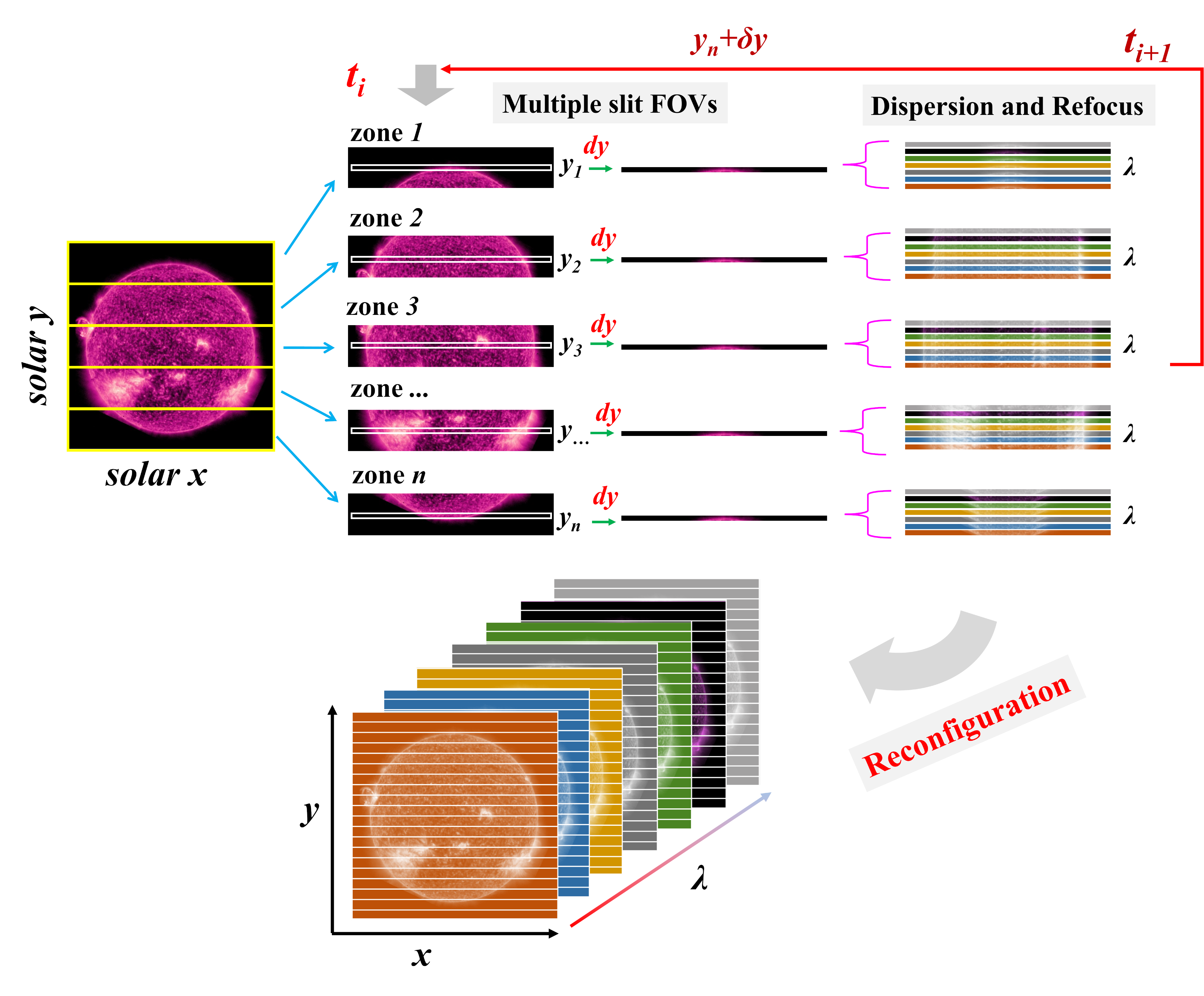}
	\caption{Overview of detection methods. The solar image in the example is from the Solar Upper Transition Region Imager (SUTRI, \citeauthor{bai2023}, \citeyear{bai2023}), detected at 17:49:12 UT on September 23$^{rd}$, 2022.}
	\label{fig:1}
\end{figure}

We use a multi-slits grating spectrograph to achieve the above detection. The whole solar disk can be divided into $n$ regions in the $y$ direction, as shown in Figure \ref{fig:1}, and $n$ distantly separated slits are then designated for each region individually. At a certain time $t_i$, slits are spaced apart in the $y_n$ position, each of which allows a slice of the solar image to pass through with a FOV $dy$. Multiple slices are dispersed and reimagined, and light of different wavelengths is recorded at different positions on the detector. Therefore, a single exposure can simultaneously obtain spatial information (along the slit) and spectral information (perpendicular to the slit) in multiple slits. At the next time $t_{i+1}$, the focused solar image is scanned (i.e., rotating the primary mirror) on the multi-slit to the next position $y_n$+$\delta y$ depended on the scanning step. By scanning step by step, the spatial and spectral information of the entire Sun is recorded sequentially. Finally, we can obtain the global coronal spatial structure as well as spectral information by reconstructing multiple image slices. In fact, the instrument has to take a comprehensive approach to obtain global coronal spatial and spectral information, improve the detected cadence and avoid overlapping. The approach is inseparable from the design of optical system for multi-slits imaging spectrograph and the optimization of narrow-band multilayer structure. \par

\section{Optical system design and performance}
\label{sec:3}
\subsection{Optical system design} 
\label{subsec:3.1}

\begin{figure}
	\centering
	\includegraphics[width=0.85\linewidth]{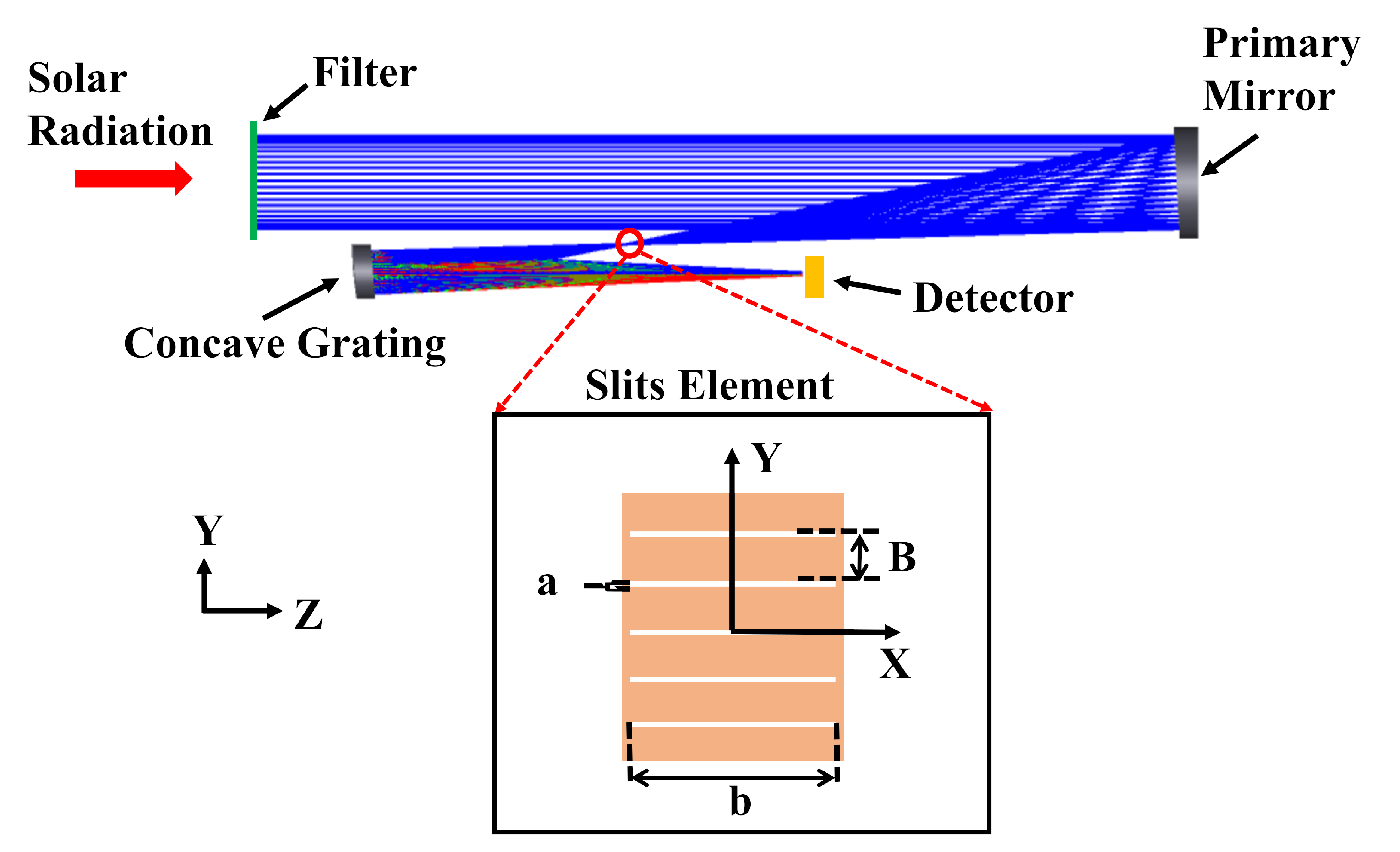}
	\caption{Schematic layout of optical system}
	\label{fig:2}
\end{figure}

The optical system of a multi-slits imaging spectrograph, as shown in Figure \ref{fig:2}, consists of an entrance filter, a primary mirror, a multi-slits element, a grating, a rear filter and a detector. The multi-slits element contains several distantly separated slits, as shown in the inset figure below, instead of a single slit as in EIS, or thirty-seven closely separated slits as in MUSE. The challenge is to determine the optimal number of slits to achieve a balanced cadence for a global FOV raster and less influence of spectral overlapping. A large number of slits can divide the entire FOV into more slices, increasing the cadence. However, this may introduce severe overlap from adjacent slits in the case of a continuous spectral range with multiple spectral lines. A small number of slits can effectively avoid overlapping, but each slit will require more time to scan a larger field of view. In addition, the transmission efficiency and bandwidth of the optical element, which depend on the materials and parameters of the multilayer mirror and grating, will also affect the overlap from adjacent slits. The optical system parameters also need to be optimized to meet the spatial resolution and spectral resolving power.\par

We first provide a general multi-slits EUV imaging spectrometer design process through a series of basic physical relationships between optical performance and system structural parameters. The spectral spacing $d\lambda$ of adjacent slits depends on the effective focal length $F$ of the optical system, FOV spacing of slits $\Delta\theta$, and the reciprocal linear dispersion $\sigma$. In order to avoid overlapping, the $\Delta\lambda$ of the observed bands should be smaller than the spectral spacing d$\lambda$ of adjacent slits (Equation (\ref{Eq.1})). The smaller the spot size of the image plane or pixel size of the detector, the better the spatial resolution $\delta$ (Equation (\ref{Eq.2})) and the spectral resolving power $R$ (Equation (\ref{Eq.3})). These relationships, as well as the specific parameters for the observations, are described as follows:

\begin{equation} \label{Eq.1}
	d\lambda = F \times \tan \frac{\Phi}{n} \times \sigma > 1.5 \, \text{nm} ,
\end{equation}

\begin{equation} \label{Eq.2}
	\delta = \arctan \frac{\max(ss,sp)}{F} \approx \frac{\max(ss,sp)}{F} \times 206265 \approx 6^{\prime\prime} ,
\end{equation}

\begin{equation} \label{Eq.3}
	R =  \frac{\lambda_0}{\Delta\lambda_0} = \frac{\lambda_0}{\max(ss,sp)\times\sigma} > 500 \, (\lambda_0 = 19.0 \, \text{nm}) .
\end{equation}
where $\sigma=\lambda_\sigma/x_\sigma$, which express the spectral range included in a unit distance on the image plane. It is a parameter used to evaluate the degree of spectral dispersion. $ss$ represents the spot size of the image plane, coupling diffraction effects and geometric aberration. $sp$ denotes the sampling period of the detector, and 206265 is the conversion factor from radians to arcseconds.\par

The performance of an optical system depends on the optical elements. The reciprocal linear dispersion, $\sigma$, is determined by the grating parameters (Equation (\ref{Eq.4})). The effective focal length of the optical system $F$ (Equation (\ref{Eq.5})) depends on the focal length of the primary mirror $f_0$ and the magnification of the grating $m_g$ (Equation (\ref{Eq.6})). These parameters can be expressed more specifically as:

\begin{equation} \label{Eq.4}
	\sigma = \frac{p\cos\beta(\lambda)\cos(\gamma)}{mr_1},
\end{equation}

\begin{equation} \label{Eq.5}
	F = f_0 \times m_g,
\end{equation}

\begin{equation} \label{Eq.6}
	m_g = \frac{r_1}{r_0}.
\end{equation}
where $p$ is the grating period, $\beta(\lambda)$ is the diffraction angle, $\gamma$ is the angle between the detector plane and the incident light, $m$ is the diffraction order, $r_1$ is the distance from the grating to the detector, and $r_0$ is the distance from the slits to the grating.\par

Considering the fabrication capability of practical EUV optical elements, some parameters are limited to a specific range. The detector adopts a 2048 $\times$ 2048 pixel array with a pixel size of 11 $\mu$m $\times$ 11 $\mu$m, and 3 pixels will be the sampling period to better satisfy the Nyquist-Shannon sampling theorem (Equation (\ref{Eq.7})). The primary mirror is an off-axis parabolic mirror with a diameter of 150 mm (Equation (\ref{Eq.8})), chosen as a compromise between the effective collecting area and off-axis aberration of the system. In addition, our detection scheme is not for extreme resolution and is a narrow observation band, so an ellipsoidal grating with uniform-line-space can meet the requirements and is preferred. The line density of the grating is further limited to no more than 3600 line $\cdot$ mm$^{-1}$, meaning the grating period $p$ is greater than 277.77 nm (Equation (\ref{Eq.9})). The grating magnification ranges between 1.5 and 2.5 (Equation (\ref{Eq.10})), which determines the slit width $a$ in combination with the pixel size. The limitations are listed as follows:

\begin{equation} \label{Eq.7}
	sp = 11 \,\text{$\mu$m} \times 3 = 33 \,\text{$\mu$m} ,
\end{equation}

\begin{equation} \label{Eq.8}
	D = 150 \,\text{mm} ,
\end{equation}

\begin{equation} \label{Eq.9}
	p > 277.77 \,\text{nm} ,
\end{equation}

\begin{equation} \label{Eq.10}
	1.5 < m_g < 2.5 ,
\end{equation}

Off-axis aberration, which seriously degrades optical performance, is another issue that needs to be overcome. Many studies have contributed to the analysis of off-axis aberrations for a EUV imaging spectrograph with concave grating \citep{Beutler1945TheTO, Harada1995DesignOH, Harada1998DesignOA, Thomas2003TVLS, Thomas2004EllipticalVL}. It is still extremely difficult to obtain the minimum aberration by solving parametric equations analytically due to its multi-parameter nonlinear coupling process. We can, nevertheless, obtain the best parameters optimized through numerical simulation methods. The relationships and constraints mentioned above are crucial references for establishing the initial parameters and optimization objectives of a numerical model. In our work, optical simulation software Zemax OpticStudio is used to reduce the aberrations and to minimize the spot size as much as possible for the wavelength of 18.3 nm, 19.0 nm and 19.8 nm. Multiple field of view points, including ($0^\prime$, $0^\prime$), ($0^\prime$, $-/+10^\prime$), ($0^\prime$, $-/+20^\prime$), ($-/+10^\prime$, $0^\prime$) and ($-/+20^\prime$, $0^\prime$), are used in the optimized merit function to achieve better resolution across all observation fields. Optimized parameters mainly include the distance from the slits to the grating $r_0$, the distance from the grating to the detector $r_1$, the grating line density, the grating surface shape parameters and the detector angle, etc. The specific parameters of the optical system and elements after optimization are shown in Table \ref{table:2}.\par

\begin{table}
	\caption{Specific parameters of the optical system and elements}      
	\label{table:2}      
	\centering   
	\begin{tabular}{c c c}   
		\hline      
		System and Elements & Parameters& Optimization Value \\   
		\hline                     
		\multirow{9}{*}{System design} & Raster FOV $\Phi$/ $\prime$ & $40^{\prime} \times 50^{\prime}$ in 200 raster steps of $3^{\prime\prime}$ with 5 slits \\
		& Wavelength $\lambda$/ nm & 18.3, 19.0 19.8 \\
		& Effective Focal Length $F$/ mm & 1533.492 \\
		& Reciprocal linear dispersion $\sigma$/ (nm $\cdot$ mm$^{-1}$) & 0.398 \\
		& Slit number $n$ & 5 \\
		& Slit space $\Delta\theta$/ $\prime$ & 10 \\
		& Slit spectral space $d\lambda$/ nm & \textbf{1.71} \\
		& Slit width $a$/ $\mu$m & 15\\
		& Slit FOV $dy$/ $\prime\prime$ & 3.43 \\
		\hline
		\multirow{5}{*}{Primary mirror parameter} & Mirror type & Off-axis parabolic mirror \\
		& Diameter $D$/ mm & 150 mm\\
		& Radius of curvature $R_0$/ mm & 1800 \\
		& Off-axis distance / mm & \textbf{100} \\
		& Conic & --1 \\
		\hline
		\multirow{9}{*}{Grating parameter} & Grating type & Ellipsoidal uniform-line-space grating \\
		& Diameter/ mm & 80 \\
		& Line density/ (line $\cdot$ mm$^{-1}$) & 3595 \\
		& Grating period $p$/ nm & 278.16 \\
		& Surface Value & a=0.001781;b=0.001779;c=605.826 \\
		& Diffraction order $m$ & --1 \\
		& Distance from the slits to the grating $r_0$/ mm & 413.80 \\
		& Distance from the grating to the detector $r_1$/ mm & 700.78 \\
		& Grating magnification $m_g$ & 1.69 \\
		\hline
		\multirow{5}{*}{Detector design} & Pixel arrays & 2048 $\times$ 2048 \\
		& Pixel size/ ($\mu$m $\times$ $\mu$m) & 11 $\times$ 11 \\
		& Spatial sampling per pixel/ $\prime\prime$ & 1.46 \\
		& Spectral sampling per pixel/ nm & 0.0043 \\
		& System spatial resolution/ $\prime\prime$ & 4.40 \\
		& System spectral resolution/ nm & 0.013 \\
		& System spectral resolving power $\lambda_0$/$\Delta\lambda_0$ & 1461($\lambda_0$ = 19.0 nm) \\
		\hline                      
	\end{tabular}
\end{table}	

\subsection{Optical performance} 
\label{subsec:3.2}

\begin{figure*}
	\centering
	\includegraphics[width=0.8\linewidth]{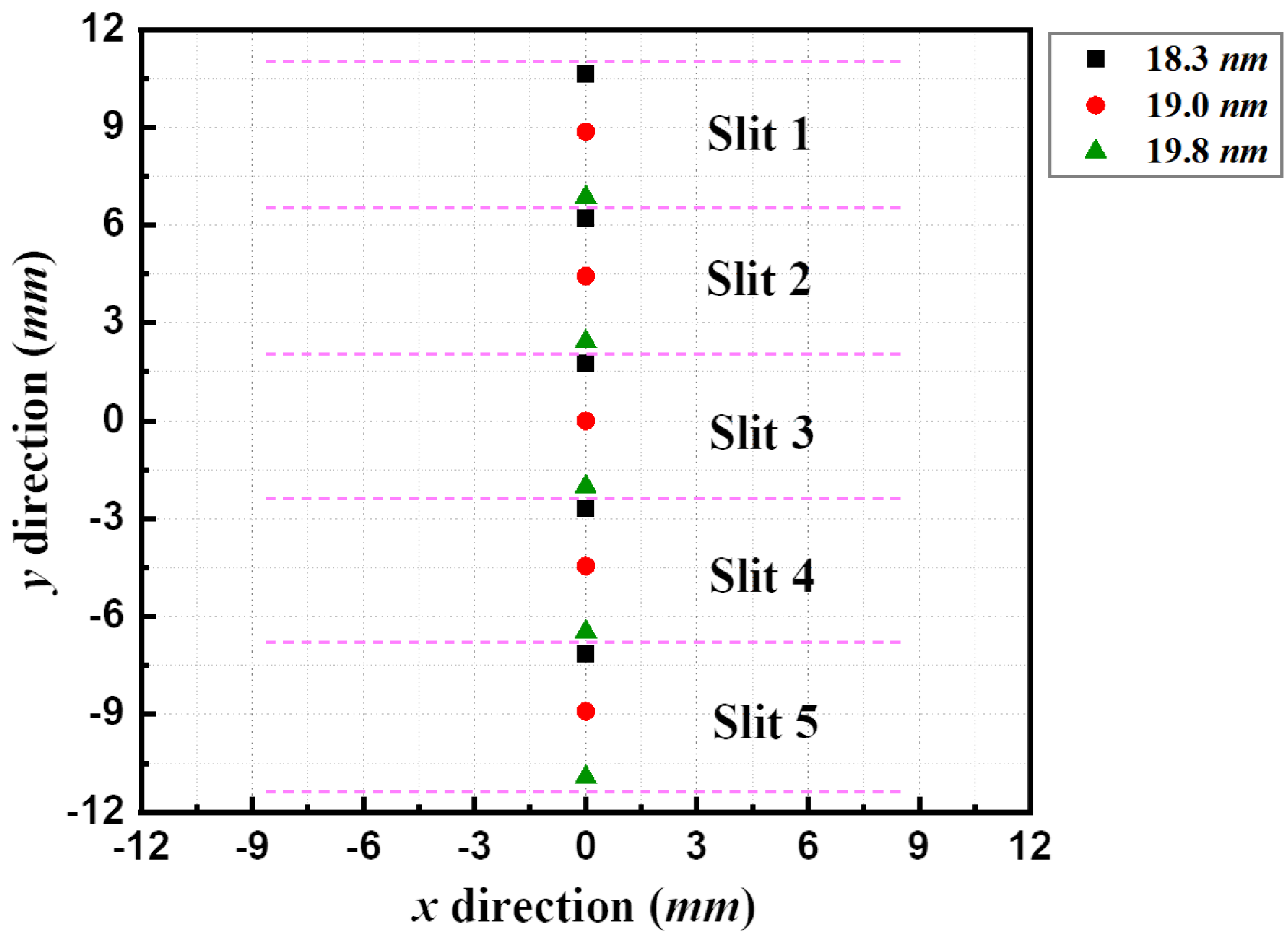}
	\caption{Position distribution of light spots of different wavelengths at the image plane}
	\label{fig:3}
\end{figure*}

In our design, the full-disk FOV is divided into five regions, i.e., the number of slits $n$ is 5, and the spacing between two nearby slits in the FOV, denoted as $\Delta\theta$, is $10^\prime$. By using ray tracing, the position distribution of light at different wavelengths from each slit FOV can be obtained on the image plane (Figure \ref{fig:3}). The five slits are positioned at $y$ = $-20^\prime$, $-10^\prime$, $0^\prime$, $10^\prime$, and $20^\prime$. For each slit, the central wavelength is 19.0 nm, and the light spots at 18.3 nm--19.8 nm are successively separated. At the same time, the light spots from the adjacent slits are not mixing with each other. For example, the 19.8 nm light spot from slit \textbf{1} does not overlap with the 18.3 nm light spot from slit \textbf{2}. The results indicate that there is no overlap from the adjacent slits for the input spectral range ($\Delta\lambda$ = 1.5 nm).\par

\begin{figure*}
	\centering
	\includegraphics[width=0.95\linewidth]{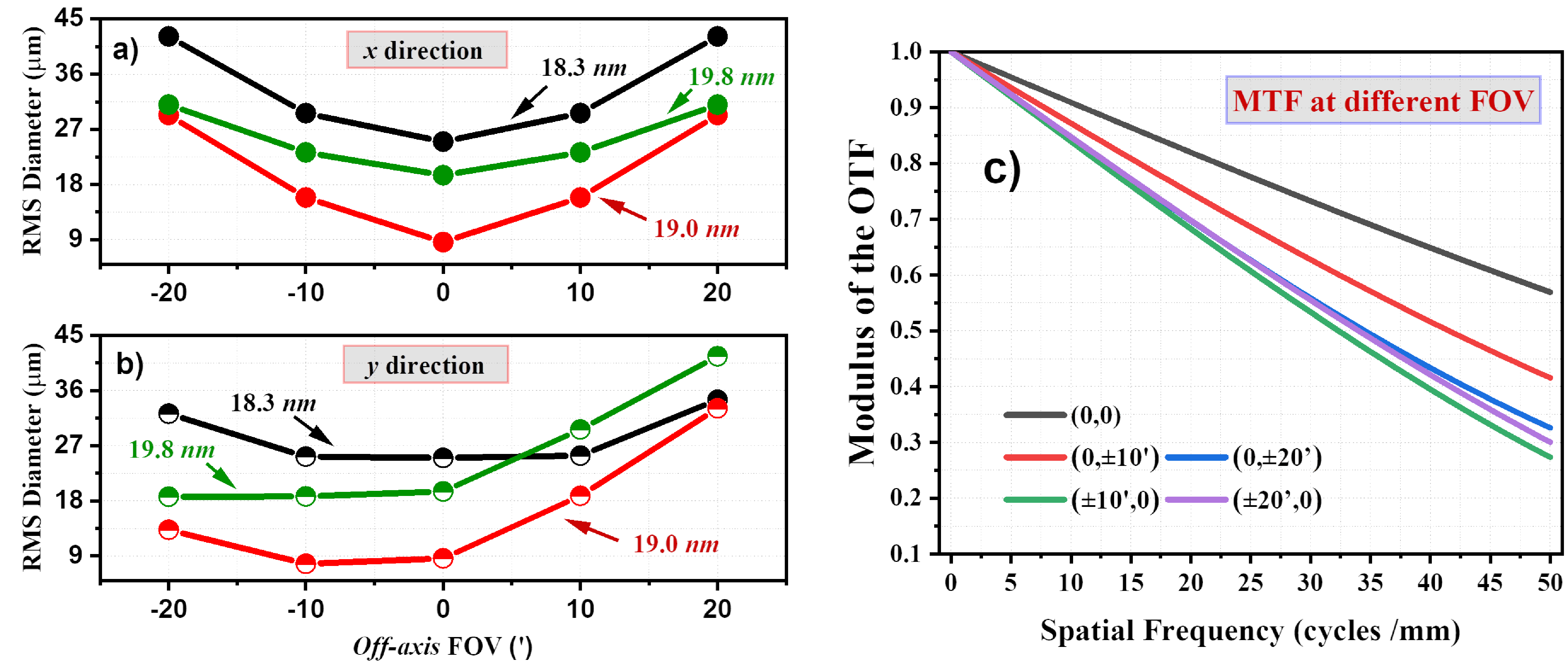}
	\caption{Performance of optical systems in different off-axis FOV: Spot RMS diameter of different wavelengths of $x$ direction (a) and $y$ direction (b) and MTF curves of the image plane (c).}
	\label{fig:4}
\end{figure*}

The Airy spot radius of the optical system is about 0.23 $\mu$m at the wavelength of 19 nm, which is much smaller than the radius caused by aberration. The spherical aberration of the grating is the main factor affecting the performance of the optical system, while the coma aberration from the primary mirror and grating is balanced with each other. In addition, there is a little pincushion type distortion in the image plane, resulting in a maximum spectral curvature of about 0.22\%. More analysis of spot size at the image plane can refer to the ray tracing results. Figure \ref{fig:4} (a) and (b) show the root mean square (RMS) spot diameter of three different wavelengths in the five off-axis FOVs along the $x$ direction and $y$ direction, respectively. The spot diameter at the center of FOV is the smallest, and it gradually increases as the deviation from the center of FOV becomes larger in either the $x$ or $y$ direction. At the wavelength of 19.0 nm, the spot diameter in the central FOV is about 8.62 $\mu$m, and the spot size of the off-axis FOV at $-/+20^\prime$ in the $x$ and $y$ directions is 29.34/29.34 $\mu$m and 13.26/33.09 $\mu$m, respectively. In the case of 18.3 nm and 19.8 nm, the tendency in spot size with off-axis FOV are consistent, but the magnitudes are greater, influenced by grating aberration. Figure \ref{fig:4}(c) presents the Modulation Transfer Function (MTF) curves of the image plane at different off-axis FOV. The curve represents the average value for each FOV at different wavelengths. At a frequency of about 30.3 cycles/mm (3 pixels per period), the MTF values for all FOVs exceed 0.52, meeting the detection requirements.\par

\begin{figure*}
	\centering
	\includegraphics[width=0.85\linewidth]{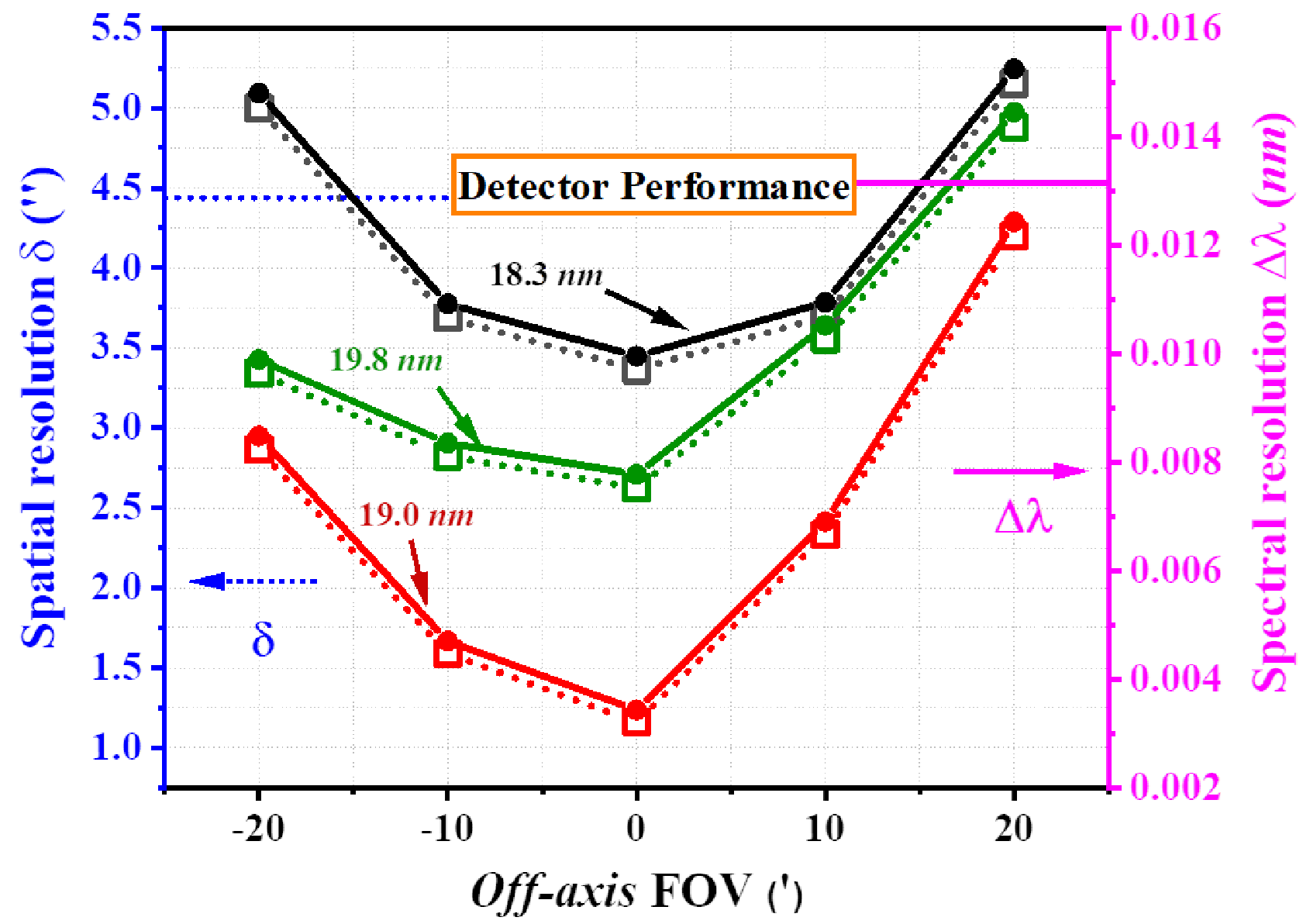}
	\caption{Spatial resolution (dot lines) and spectral resolution (solid lines) of different wavelengths in different off-axis FOV, detector performance based on the sampling period also marked with blue dots and purple solid lines}
	\label{fig:5}
\end{figure*}

Figure \ref{fig:5} illustrates the spatial and spectral resolution of the three different wavelengths in the five off-axis FOVs, calculated using Eq.(3) and averaging the $x$ and $y$ directions. The results show that both the spatial and spectral resolution decrease with the deviation from the center FOV. The resolution corresponding to three pixels sampling with the detector is also indicated in Figure \ref{fig:5}. In other words, the resolution of the optical system is limited by the three pixel sampling. The spatial resolution of the optical system is about $4.4^{\prime\prime}$, and the spectral resolution is about 0.013 nm (i.e., the spectral resolving power exceeds 1461). The worst spatial and spectral resolutions are $5.1^{\prime\prime}$ and 0.015 nm (i.e., the spectral resolving power is greater than 1266), respectively. The spatial resolution of the scanning direction also needs to consider the slit width and scanning step. Our scheme has a slit width of $3.43^{\prime\prime}$, which means the spatial resolution in the scanning direction is about $6.86^{\prime\prime}$.\par

\subsection{Effective Areas} 
\label{subsec:3.3}

The effective area is another key parameter of the optical system that determines the number of received photons or the signal-to-noise ratio (S/N) for a single exposure. On the one hand, the bandwidth of the effective area needs to be large enough to cover the observation band. On the other hand, the sideband of effective area needs to suppress to avoid overlapping caused by excess radiation. The effective area $e(\lambda)$ of the optical system depends on the transmission efficiency of each optical element at different wavelengths, which can be expressed as:

\begin{equation} \label{Eq.11}
	e(\lambda) = A \times T_{ff}(\lambda) \times R(\lambda) \times T_{rf}(\lambda) \times E(\lambda)\times Q(\lambda).
\end{equation}
where $A$ is the effective light collecting area of the optical system, $T_{ff} (\lambda)$ and $T_{rf} (\lambda)$ are the transmissions of the front and rear filters, respectively. $R(\lambda)$ represents the reflectance of the primary mirror coated with multilayers. $E(\lambda)$ is the diffraction efficiency of multilayer gratings, which includes both the groove efficiency and the reflectivity of the multilayer coatings, and $Q(\lambda)$ is the detector quantum efficiency.\par

\begin{figure}
	\centering
	\includegraphics[width=0.95\linewidth]{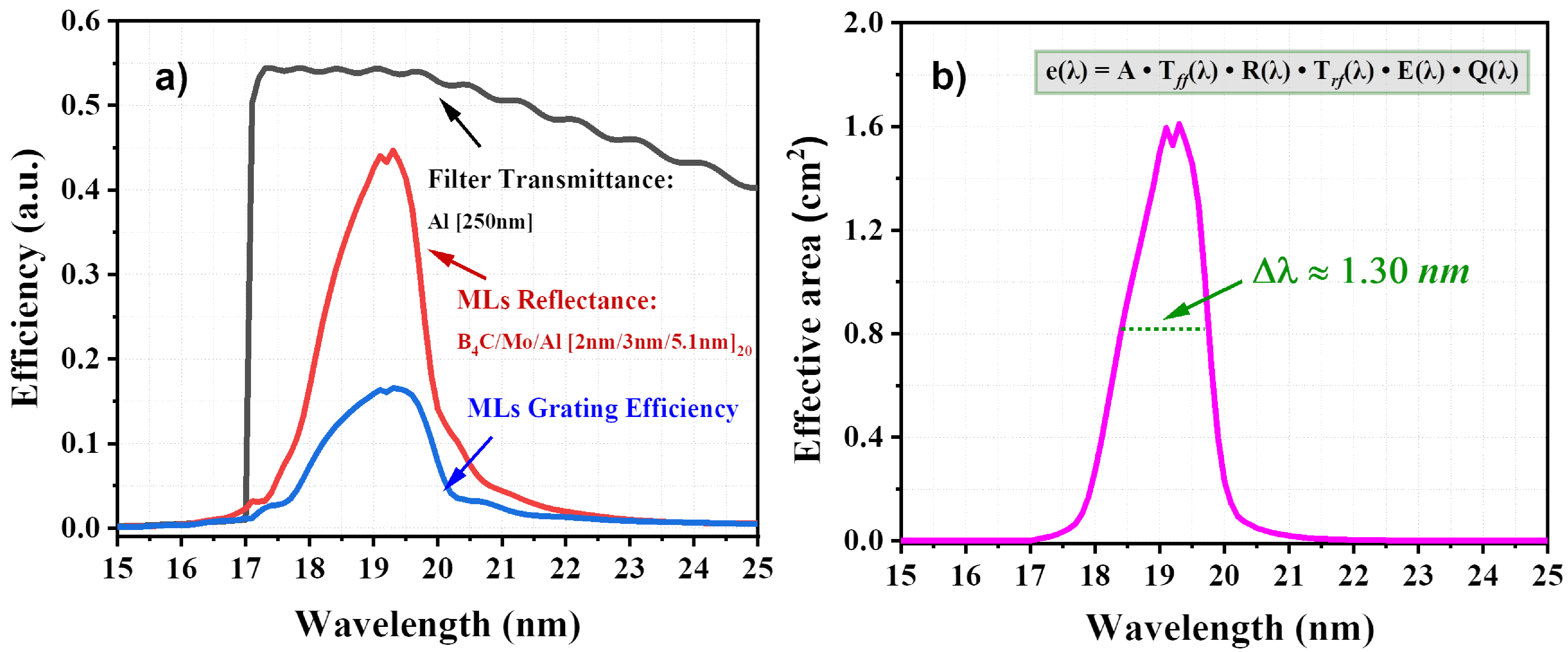}
	\caption{Theoretical calculation efficiency curve of elements (a) and the effective area of the optical system (b).}
	\label{fig:6}
\end{figure}

The material of the front filter and the spectrometer entrance filter is Al thin-films, which have high transmittance in the band of 17.1--25.0 nm and have been widely used in many EUV solar instruments such as EIS \citep{Korendyke2006OpticsAM} and Hi-C \citep{Rachmeler2019TheHC}. The design thickness of both Al filters is 250 nm, while 5 nm of aluminum oxide is applied on both sides to adapt to the actual situation. The transmittance curve of the filters is calculated using IMD software \citep{David1998IMD}, as shown in Figure \ref{fig:6}(a). The results show that the transmittance in the 18.3--19.8 nm band is greater than 53\%.\par

The multilayer structure is designed for the primary mirror to enhance the transmission efficiency of of the target waveband. In the 18--20 nm band, an Al and B$_4$C material combination is selected, and a thin Mo layer is added to improve the interface roughness \citep{Delmotte2013DevelopmentOM, Corso2019ExtremeUM}. The structural parameters of Al/Mo/B$_4$C multilayers, including d-spacing, the number of multilayer $N$, etc., are optimized using IMD software to achieve high reflectivity in the observation band while reducing the sideband. The optimized multilayer structure is Al/Mo/B$_4$C [5.1 nm/3 nm/2 nm], and the multilayer number $N$ is 15. Additionally, a B$_4$C protective layer with a thickness of 10 nm is added on top to prevent oxidation and ion bombardment from solar wind. The reflectance curve at the incident angle of $3.17^\circ$ is shown in Figure \ref{fig:6}(a), where the interface roughness of the calculation model is set to 0.6 nm. The results indicate that the reflectance at 19.3 nm is about 43.5\%, while the values are 27.9\% and 27.2\% at the wavelengths of 18.3 nm and 19.8 nm, respectively.\par

The optimized multilayer structure is also combined with the grating. Based on the incident angle and multilayer structure, the grating groove depth is determined to be 4.76 nm, and the duty cycle is 0.5. Figure \ref{fig:6}(a) shows the diffraction efficiency curves of the $-1^{st}$ order of the multilayer grating as a function of wavelength. These curves are calculated numerically using rigorous coupled wave analysis (RCWA). The results indicate that the diffraction efficiency of the multilayer grating at a wavelength of 19.3 nm is approximately 16.5\%.\par

Finally, the effective area $e(\lambda)$ of the whole system can be calculated using Equation (\ref{Eq.11}) with the efficiency values of the elements obtained above. The result is shown in Figure \ref{fig:6}(b). The quantum efficiency of the detector is set at 0.5, while the reduction factor of the actual multilayer grating efficiency is 0.85. This reduction factor accounts for the interface roughness of the multilayers and the variation in grating groove, compared to the ideal calculation model. The effective area at 19.3 nm is about 1.61 cm$^2$, which is 5 times larger than that of EIS. The values at 18.3 nm and 19.8 nm are both approximately half of the peak (0.80 cm$^2$), while the effective areas are reduced to 0.035 cm$^2$ and 0.049 cm$^2$ at the wavelengths of 17.5 nm and 20.5 nm, respectively. The small overlaps from the adjacent slits could be separated algorithmically.\par

\section{\textbf{End-to-end optical response and photon budget}}
\label{sec:4}

It is critically important to evaluate the instrument response to solar radiation to better understand its sensitivity under various measurement conditions. Using the synthetic spectra and the effective areas, the photon numbers registered in each detector pixel per second $N(\lambda)$ (photons s$^{-1}$ pixel$^{-1}$) can be expressed as:

\begin{equation} \label{Eq.12}
	N(\lambda) = I(\lambda) \times e(\lambda) \times \delta _{sr} \times \delta _{\lambda}
\end{equation}
where $I(\lambda)$ (photons cm$^{-2}$ s$^{-1}$ sr$^{-1}$ \AA $^{-1}$) is the intensity of the solar radiation calculated with CHIANTI \citep{Dere1997CHIANTIA, Young2002CHIANTIAnAD}, $e(\lambda)$ is the effective area including all factors from Equation (\ref{Eq.11}), and $\delta_{sr}$ and $\delta_\lambda$ represent the solid angle and spectral resolution per pixel, respectively.\par

\begin{table}
	\caption{Calculated solar radiation and detection photon numbers of the instrument}      
	\label{table:3}      
	\centering 
	\begin{tabular}{c c c c c}  
		\hline    
		Ion and Wavelength & \multicolumn{2}{c}{Solar radiation} & \multicolumn{2}{c}{Detected photon number} \\  
		\AA & \multicolumn{2}{c}{$\times$10$^{12} \,$photons cm$^{-2}$ s$^{-1}$ sr$^{-1}$ \AA $^{-1}$ } & \multicolumn{2}{c}{photons s$^{-1}$ pixel$^{-1}$} \\
		\hline{2-5}
		{} & AR & QS & AR & QS \\
		\hline                  
		Fe~{\sc{viii}} 185.21 \AA & 34.87 & 5.14 & 73.0 & 10.78 \\
		Fe~{\sc{x}} 184.54 \AA & 54.17 & 4.26 & 98.5 & 7.7 \\
		Fe~{\sc{xi}} 188.22 \AA & 68.80 & 4.09 & 195.5 & 11.6 \\
		Fe~{\sc{xii}} 186.89 \AA & 88.74 & 4.11 & 207.9 & 9.6 \\
		Fe~{\sc{xii}} 193.51 \AA & 116.92 & 4.37 & 424.0 & 15.8 \\
		Fe~{\sc{xii}} 195.12 \AA & 174.88 & 8.02 & 573.9 & 26.3 \\
		\hline                      
	\end{tabular}
	\footnotetext{\textbf{Note.} The calculation of solar radiation uses two standard CHIANTI differential emission measures (AR and QS), assuming an electron density of 10$^9$ cm$^{-3}$. We also assume ionization equilibrium and adopt the coronal abundance from \citeauthor{Schmelz2012composition} \citeyear{Schmelz2012composition}. The spectral line shape adopts Gaussian profile and the line width of about 0.2 \AA is given by a combination of thermal broadening and instrumental broadening.}
\end{table}	

Table \ref{table:3} presents the input solar radiation and received photon numbers per pixel of six typical lines (Fe~{\sc{viii}} 185.21 \AA, Fe~{\sc{x}} 184.54 \AA, Fe~{\sc{xi}} 188.22 \AA, Fe~{\sc{xii}} 186.89 \AA, Fe~{\sc{xii}} 193.51 \AA, and Fe~{\sc{xii}} 195.12 \AA) in AR and QS. We set a single detected time of 1.5 s (including exposure time of 1.3 s and raster scan time of 0.2 s by rotating the primary mirror) and a raster step of $3^{\prime\prime}$. To scan the whole FOV with 5 slits, we need ($3000^{\prime\prime}$/5)/$3^{\prime\prime}$ = 200 scanning steps and complete the scan of the whole FOV within 200$\times$ 1.5 s = 300 s. For the detection in AR, the typical photon numbers are greater than 95 photons s$^{-1}$ pixel$^{-1}$ with a 1.3 s exposure time for all spectral lines. For the detection in QS, the typical photon numbers of the strong radiation lines (Fe~{\sc{xii}} 193.51 \AA, Fe~{\sc{xii}} 195.12 \AA, etc.) are within 20--32 photons s$^{-1}$ pixel$^{-1}$ for a single exposure time. While for the detection of weak radiation lines (Fe~{\sc{viii}} 185.21 \AA, Fe~{\sc{xi}} 188.22 \AA, etc.) in QS, we could sacrifice the temporal resolution to increase the S/N ratio. Either the multi-frames integration methods or setting a longer exposure time can be adopted. For the regions with flares, the exposure time should be set much lower to avoid saturation. \par

\begin{figure}
	\centering
	\includegraphics[width=0.99\linewidth]{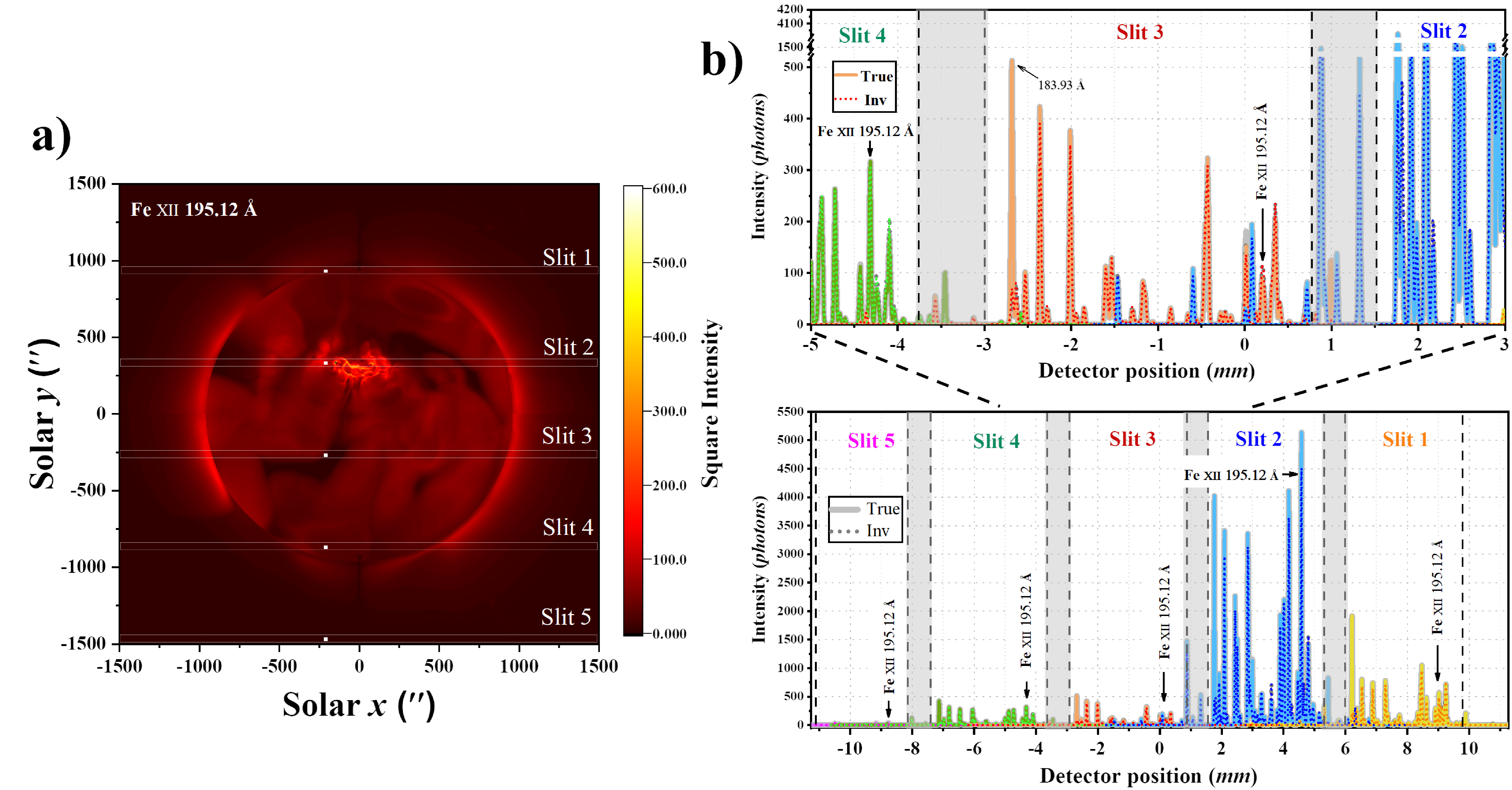}
	\caption{(a) shows the synthesized global intensity map for Fe~{\sc{xii}} 195.12 \AA. Bottom panel of (b) shows total spectra from 5 slits marked with white squares in Figure 7(a) and top panel of (b) shows the zoom-in spectrum from slits 2 to 4. For improved clarity, two breakpoints (the first one ranges from 520 to 1400, and the second one ranges from 1600 to 4000) have been set for the ordinate of top panel in (b).}
	\label{fig:7}
\end{figure}

Although we avoid overlapping in the design of optical system for the input spectral range ($\Delta\lambda$ = 1.5 nm), there is still small overlapping due to the unavoidable sideband of the effective area. A decomposition method is employed to extract spectra from each slit in the observed spectra and generate the global coronal map. The method, which adheres to the basic inversion framework outlined by \citeauthor{Cheung2015THERMALDW} (\citeyear{Cheung2015THERMALDW, Cheung2019MulticomponentDO}), has been evaluated for a multi-slit extreme ultraviolet imaging spectrograph by \citeauthor{chen2024global} \citeyear{chen2024global}. \citeauthor{chen2024global} \citeyear{chen2024global} utilized a numerical model (from Predictive Science Inc.) as the observation spectrum (ground truth) to examine the influence of spectra overlapping. The results presented here are achieved by executing the program in combination with the parameters of the instrument. The noticable difference between the two schemes is slit separation on the plane of sky and spectral spacing of the inter-slits on the spectrogram. We optimized the spectral spacing to 17.1 \AA \ to avoid overlapping from adjacent slits as much as possible, while Chan's scheme had a inter-slit spacing of 1 \AA \ mainly to verify the decomposition method. In Figure \ref{fig:7}(a), we synthesized the global intensity map of the Fe~{\sc{xii}} 195.12 \AA. The structures of coronal loops from AR and CHs are clearly resolved. The bottom panel of Figure \ref{fig:7}(b) displays the synthesized (solid lines) and inverted spectra (dot lines) from five slit pieces marked with white squares in Figure \ref{fig:7}(a), and the top panel showcases the zoom-in spectrum from slits \textbf{2} to \textbf{4}. Notably, in Figure \ref{fig:7}(b), there is an inevitable overlap of spectra from adjacent slits, denoted by a gray bar. Due to significant variations in radiation intensity, several lines from slit \textbf{2} (located near the AR) exhibit comparable strength to the primary lines from slit \textbf{3} (located near the coronal hole). Limited by computing resources, the inversion here primary prioritizes the acquisition of accurate information for six primary lines. Other lines, such as 183.93 \AA \ on the left edge of slit \textbf{3} (at the detector position of about --2.67 mm), may not include potential parameters during the inversion process, resulting in large deviation. One possible solution in the future would be to extend our parameter space of the response matrix to obtain more accurate information, while this would result in increased computation time.\par

\section{Conclusions}
\label{sec:5}
The paper describes a method to optimize multi-slit extreme ultraviolet imaging spectrograph for global coronal plasma diagnostic with high cadence. Five narrow slits are designed to divide the entire solar field of view, which can increase the cadence by 5 times relative to a single slit and raster scan full-disk with 5 minutes. Considering the fabrication capability of practical EUV optical elements and the scientific requirements, the parameters are optimized in the spectral range of 18.3--19.8 nm with an aperture diameter of 150 mm. The spatial resolution of optical system is better than $4.4^{\prime\prime}$ and the spectral resolving power exceeding 1461. The Al/Mo/B$_4$C multilayer structure is optimized to maximize the efficiency of the reflector and grating in the observation band, resulting a peak effective area of about 1.60 cm$^2$ at 19.3 nm. Finally, the instrument performance is evaluated by calculating the photon numbers and inverting the spectra to obtain a global coronal map. This work offers an innovative approach for full-disk corona spectral diagnostics. The method is also useful for the design of a multi-slit imaging spectrograph worked at the other EUV wavelength and estimating its performance.\par

In order to realize our detection scheme, some practical challenges would be further explored in the future. First, it is crucial that the groove shaped replication growth of multilayer on ellipsoidal gratings with a line density of about 3600 line $\cdot$ mm$^{-1}$. There have been many studies on the improvement of coating process of multilayer gratings \citep[e.g.,][]{Voronov2012MoSi, Feng2021MoSi}. Recently, \citeauthor{Mahmoud2022AlMoSiC} \citeyear{Mahmoud2022AlMoSiC} reported the preparation and characterization of Al/Mo/SiC multilayer lamellar gratings with line densities of 3600 line $\cdot$ mm$^{-1}$. The results indicate that the diffraction efficiency decreased for a large multilayer number due to the deviation of grating profile. Therefore, in the future, some measures, including multilayer design and coating process improvement, are key to achieve high efficiency ellipsoidal gratings with high line density. Then high precision image stabilization system and scanning mechanism are also the key to realize full-disk FOV detection by raster scan. The slit FOV in our scheme is about $3.43^{\prime\prime}$, which means that the accuracy of the image stabilization system and the error of scanning step should be better than $0.3^{\prime\prime}$. The raster scan time of the instrument is about 0.2 s, resulting in the response time of the image stabilization system and scanning mechanism should be less than 100 ms. Finally, it is a very complicated process to reconstruct full-disk FOV spectral imaging information with many slit scanning fragments. Spectral extraction is the result of evaluating overlapping and decomposition method. Global coronal imaging is the coupling of decomposition methods with slit width, scan step, etc. \par

\bmhead{Acknowledgements}

This work was supported by National Key R\&D Program of China No. 2021YFA1600500, National Natural Science Foundation of China No. 12303088. CHIANTI is a collaborative project involving George Mason University, the University of Michigan (USA), University of Cambridge (UK) and NASA Goddard Space Flight Center (USA).

\section*{Declarations}

\begin{itemize}
\item Funding 
This work was supported by National Key R\&D Program of China No. 2021YFA1600500, National Natural Science Foundation of China No. 12303088.
\item Competing interests 
The authors declare no competing interests.
\item Ethics approval and consent to participate  
Not applicable.
\item Consent for publication 
The authors confirm that this article has not been previously published in any journal.
\item Data availability 
Not applicable.
\item Materials availability 
Not applicable.
\item Code availability  
Not applicable.
\item Author contribution 
Conceptualization: Yufei Feng, Xianyong Bai, Hui Tian.
Methodology: Yufei Feng, Xianyong Bai, Hui Tian.
Formal analysis and investigation: Sifan Guo, Lami Chan, Qi Yang, Wei Duan.
Writing - original draft preparation: Yufei Feng
Writing - review and editing: Xiaoming Zhu, Xiao Yang, Zhiwei Feng, and Zhiyong Zhang.
Supervision: Hui Tian, Yuanyong Deng.
\end{itemize}

\noindent

\bigskip

\bibliography{reference}

\end{document}